# Control of the Exciton Transfer in Quantum Dots by the Stark Effect

P. A. Golovinski [1,2]

[1] *Moscow Institute of Physics and Technology (State University), 141700 Moscow, Russia*
[2] *Voronezh State University of Architecture and Civil Engineering, 394006 Voronezh, Russia*

Resonant transfer of energy between excited states in a system of two semiconductor quantum dots is studied theoretically. The model Hamiltonian has been formulated, which allowed describe the impact on the dynamics of the resonant laser pulse, the Coulomb interaction, the static Stark effect, and the relaxation of the exciton states. Examples of calculations for efficiency of the energy transfer under different excitation conditions are presented. The control of the process by the level shift in a constant electric field is demonstrated.

The full version is submitted to the Semiconductors (in Russian)

### 1. Introduction

The physical implementation of quantum computing formulates very high demands to hardware. Such increased characteristics in compare with classical computers are caused by coherent nature of quantum computation [1]. Different realizations based on quantum optics, atomic or molecular systems are not technological and have a size much large in compare with a tipical size of the modern micro electronic devices. So, the solid state realization seems a logical way to combine the newest achievements of the super fast optoelectronics and the nano structures technology. Quantum dots with Coulomb inter dots interaction pretend to be adequate physical bases of quantum computers [1, 2]. In this scheme, quantum dot is excited by a short optical pulse and then excitation is transfered to another quantum dot due to the Förster interaction [3, 4] without electric charge transport. Up to date theoretical description of such processes does not take into account impulse character of laser excitation that is very essential for validity of the model and accurate estimation of efficiency and a time of switching of all-optical gates.

In quantum dots with different size and composition, the excitonic states have different energies also. The positions of energy levels may be controlled with the help of dc Stark effect. In following, we discuss the lowest levels of excitations in the neighbor quantum dots. The barrier between quantum dots is assumed as enough large to prevent the tunnel transition of electron from dot to dot. In the case of nearly degenerate energy levels, the external electric field splits the energy as the result of the Stark effect. It is the near field interaction of the nano size particles [5]. The paper is devoted to the formulation of the model for the resonant exciton transfer between quantum dots that are stimulated by a short optical pulse and controlled by the Stark effect gate.

## 2. The model Hamiltonian and basic equations

The Förster energy transfer in semiconductors can be analyzed in the framework of the model of two resonant states [6]. The process of the energy transfer crucially depends on the relation between the dipole-dipole interaction and the energy width. When Coulomb interaction dominates, the energy transfer has coherent manifestations, and one has a new state as a superposition of subsystems states. In the case of large relaxation in the acceptor, the dissipation is dominated in the acceptor dot too, and it is the situation, when the classical Förster model is realized.

For detailed description of the process, the different Hamiltonians are utilized. It is essential incorporate into the model the dynamics of quantum dot excitation by a short laser pulse, the energy shift by the Stark effect, and the spontaneous decay of excited states due to the emission of photons and phonons. The energy levels of the systems are depicted in Fig. 1.

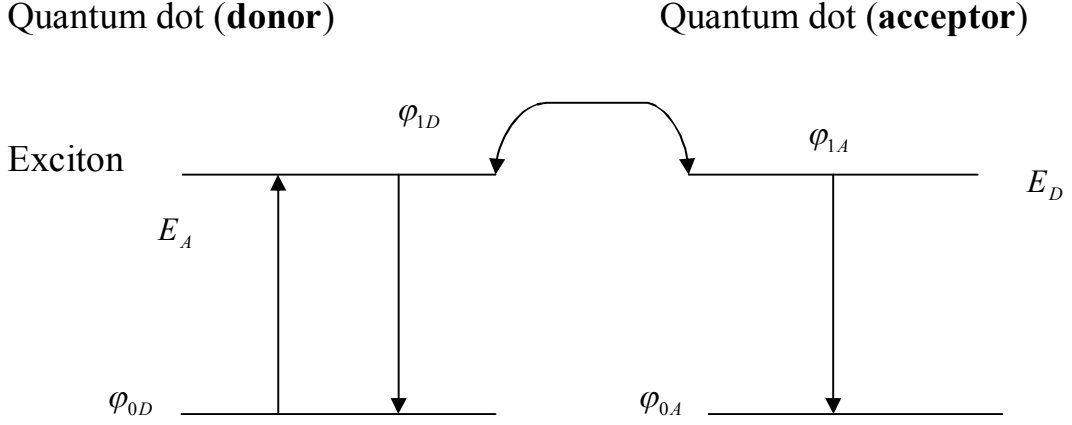

Fig. 1. The scheme of the energy levels for two interacting quantum dots.

The steady states of the combined system of two quantum dots are $|00\rangle = \varphi_{0D}\varphi_{0A}$, $|10\rangle = \varphi_{1D}\varphi_{0A}$, $|01\rangle = \varphi_{0D}\varphi_{1A}$, and $|10\rangle = \varphi_{1D}\varphi_{0A}$, $|11\rangle = \varphi_{1D}\varphi_{1A}$ were $\varphi_{0D}$ is the ground state of the donor quantum dot, $\varphi_{0A}$ is the ground state of the acceptor quantum dot, $\varphi_{1D}$ is the exciton state of the donor quantum dot, $\varphi_{1A}$ is the exciton state of the acceptor quantum dot. In this bases the wave function of the system is of the form

$$\psi = (|00\rangle, |10\rangle, |01\rangle, |11\rangle)^T = (\psi_1, \psi_2, \psi_3, \psi_4)^T \quad (1)$$

The effective Hamiltonian of two interacted quantum dots may be written as

$$H = H_0 + H_{int}(t) \quad (2)$$

where its diagonal part

$$H_0 = \begin{pmatrix} E_0 & 0 & 0 & 0 \\ 0 & \widetilde{E}_D(t) & 0 & 0 \\ 0 & 0 & \widetilde{E}_A(t) & 0 \\ 0 & 0 & 0 & \widetilde{E}_D(t) + \widetilde{E}_A(t) + V_b \end{pmatrix} \quad (3)$$

corresponds to the separated quantum dots in the field of electrostatic gate, and the interaction operator

$$H_{int}(t) = \begin{pmatrix} 0 & V_D(t) & V_A(t) & 0 \\ V_D(t) & 0 & V_F & V_A(t) \\ V_A(t) & V_F & 0 & V_D(t) \\ 0 & V_A(t) & V_D(t) & 0 \end{pmatrix} \quad (4)$$

takes into account the Förster perturbation and the laser field action. Here $E_0$ is the ground state energy of the system, $\widetilde{E}_D = E_D + \Delta E_D(t) - i\gamma_D/2$, $E_D$ is the energy of the system with exciton in the donor quantum dot; $\widetilde{E}_A = E_A + \Delta E_A(t) - i\gamma_A/2$, $E_A$ is the energy of the system with exciton in the acceptor quantum dot. The interaction of quantum dots with a laser pulse is described by operator $V(t)$, and $V_F$ is the Förster interaction, $\gamma_D$ is the spontaneous width of the donor exciton state, $\gamma_A$ is the spontaneous width of the acceptor exciton state. The Stark shift of the donor exciton state is $\Delta E_D(t)$, and the Stark shift of the acceptor exciton state is $\Delta E_A(t)$. Both energies change slowly with time. $V_b$ is the shift of the biexciton energy level by Coulomb interaction.

The Stark shift for quantum dots depends on the size and components [7, 8], that allows to control excitation with the resonant energy transfer. The experimental parameters for the quantum dots of InGa/GsaAs one may find in [9]. For quantum dot with the size $R = 5$ nm one has: $E_D - E_0 = 1.3$ eV, $|\langle \mathbf{r} \rangle| = 6 \, \overset{\circ}{}$, $\gamma = 1.04 \cdot 10^9 \, \text{s}^{-1}$ ($\tau_{decay} = 964$ ps), $V_F = 0.69$ meV ($\tau_F = 0.956$) ps, and the Stark shift $\varepsilon = 2V_F$ for electric field $F_0 = 7.547 \cdot 10^6$ V/m.

### 3. The excitation transfer induced by laser pulse

For laser excitation with a picosecond duration and comparably small relaxation times the control by the Stark shift is, in fact, equivalent to the constant detuning $\varepsilon$ of the resonance. The diagonal elements of the Hamiltonian for the case $\tilde{E}_D = \text{const}$, $\tilde{E}_A = \text{const}$. The next essential circumstance is a quasi harmonic structure of the picosecond laser pulses. It means, one can define such pulses as a product of the harmonic function of carrier oscillations

$$F(t) = f(t)\cos(\omega t) = \frac{f(t)}{2} e^{-i\omega t} + c.c. \quad (5)$$

and the envelope Gaussian function

$$f(t) = f_0 \exp(-t^2/\tau^2) \quad (6)$$

In the frame of this parameterization, the pulse duration is defined as $T = \sqrt{2}\tau$. The frequency of the carrier wave we assume is nearly equal to the excitation energy for the donor quantum dot: $\omega \approx \tilde{E}_D - E_0$.

After substitution $\psi_i(t) = a_i(t)\exp(-iE_i t)$ to the Shrödinger equation

$$i\frac{\partial \psi}{\partial t} = H\psi \quad (7)$$

one comes to interaction representation ($\hbar = 1$)

$$i\dot{a}_f(t) = \sum_i \langle f|H_{int}(t)|i\rangle \exp(i\omega_{fi}t)a_i(t),$$
$$\omega_{fi} = E_f - E_i \quad (8)$$

In the rotation wave approximation one obtains

$$H_{int}(t) = \begin{pmatrix} 0 & W(t)e^{i(\omega_{21}-\omega)t} & 0 & 0 \\ W(t)e^{-i(\omega_{21}-\omega)t} & -i\gamma_D/2 & V_F e^{i\omega_{32}t} & 0 \\ 0 & V_F e^{-i\omega_{32}t} & -i\gamma_A/2 & 0 \\ 0 & 0 & 0 & 0 \end{pmatrix},$$ (9)

$$W(t) = (\mathbf{r}_{12}\mathbf{e})\frac{f(t)}{2}$$

This representation eliminates all "rapid variables" with conservation of slowly changing coefficients, and the system became of three levels. It contains the bound state of the system of two quantum dots and the excited states of separated quantum dots. In Fig. 2 the dynamics of donor and acceptor population is depicted as a function of time. Fig. 2.a shows the dynamics for detuning of the resonance that is equal to 2 meV. The taken parameters are $V_F = 0.1$ meV, $\gamma = 10^{-2}$ meV [10], the magnitude of laser pulse perturbation is $W_0 = (\mathbf{r}_{12}\mathbf{e})f_0/2 = 50$ meV and the pulse duration is $T = 4$ ps.

a)

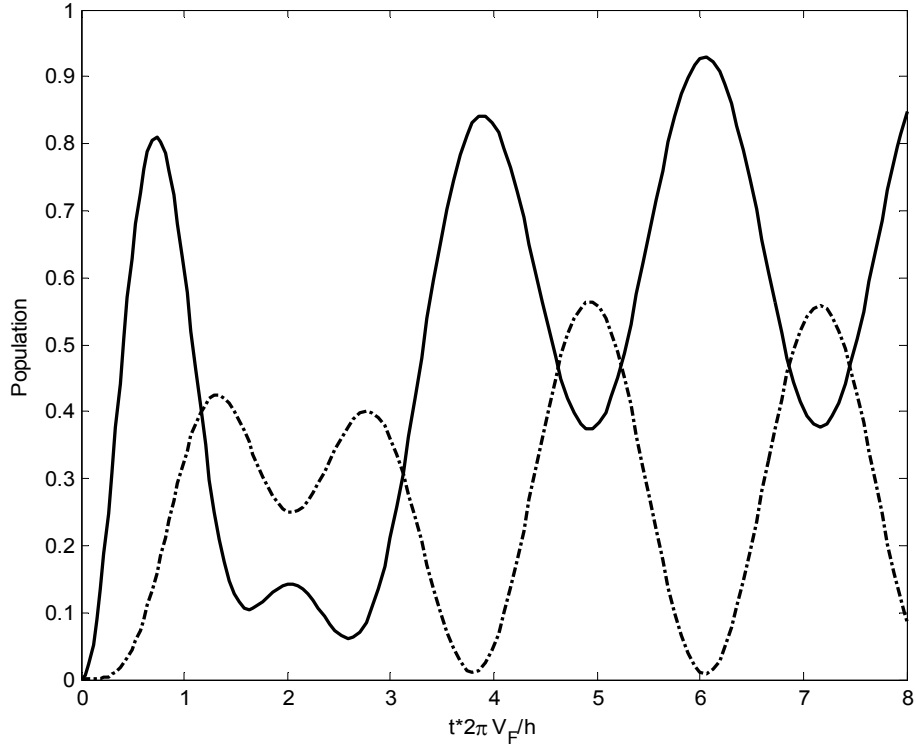

b)

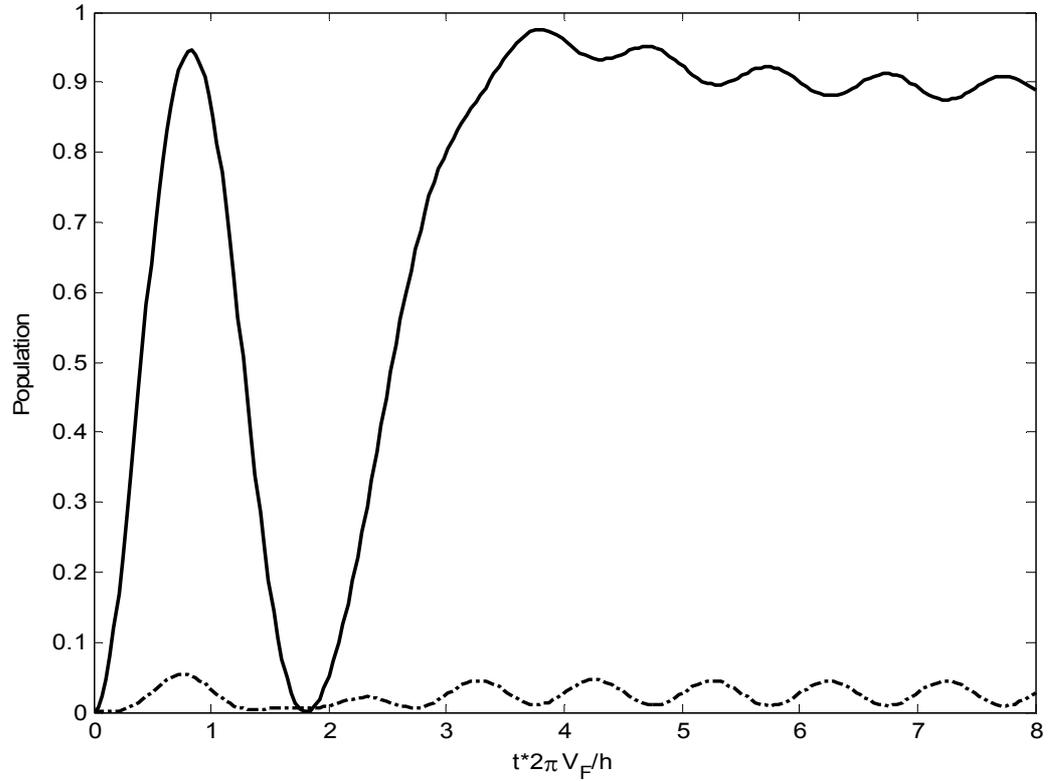

Fig. 2. Variation of the population for excited donor and acceptor states according to the different detunings: a) – meV; b) – meV. Solid line corresponds to the donor state, dot-dashed line is the same, for the acceptor.

Very fast the excitation of the donor state is accompanied by transfer energy to acceptor and by periodical oscillations of populations in both subsystems. The power of excitation gradually decreases because of the relaxation in quantum dots. We took the optimal laser pulse duration for excitation and transfer energy from the donor to the acceptor. In Fig. 2.b the dynamics is shown for the energy detuning 6 meV. For this case of relatively moderate detuning, the energy transfer is suppressed. The results of numerical simulation demonstrate strong dependence of the efficiency of the process on the relative Stark shift of the levels in donor and acceptor quantum dots. The transfer coefficient may be more than 0.5. In the considered example the characteristic time of excitation with the energy transfer is equal ~5 ps.

## 4. Conclusions

In this work we theoretically investigated the dynamics of the exciton transition between two quantum dots, that is controlled by the Stark effect. The process is stimulated with the help of a short laser pulse. For computer simulation we utilized the model Hamiltonian, describing the laser excitation and the resonant Förster energy transfer in the presence of static electric field. The system of two quantum dots has a good prospects as a candidate to the picoseconds switch.